\documentstyle[12pt]{article}
\newcounter{eqn}
\def\lab{\refstepcounter{eqn}\eqno(\arabic{eqn})}
\def\l#1{\lab\label{#1}}
\def\r#1{(\ref{#1})}

\begin{document}
\begin{titlepage}
\title{RENORMALIZATION OF THE SEMICLASSICAL HAMILTONIAN FIELD THEORY}

\author{
 V.P.Maslov and  O.Yu. Shvedov
\\
{\small {\em Chair of Quantum Statistics and Field Theory,}}\\
{\small{\em Department of Physics, Moscow State University }}\\
{\small{\em Vorobievy gory, Moscow 119899, Russia}} }

\end{titlepage}
\maketitle

\begin{flushright}
hep-th/9807061 \\
Talk presented at the \\
International Seminar ``Quarks-98'',\\
Suzdal, Russia, May 17-24, 1998
\end{flushright}

\footnotetext{e-mail: olshv@ms2.inr.ac.ru}

\begin{center}
{\bf Abstract}
\end{center}

\begin{flushright}

\parbox{17cm}{
The Hamiltonian approach to the quantum field  theory  is  considered.
Since there  are  additional difficulties such as the Haag theorem and
Stueckelberg divergences,  renormalization   of   the   time-dependent
dynamical quantum   field   theory   is  much  more  complicated  than
renormalization of the S-matrix.  It  is  necessary  to  consider  the
regularized theory  with  ultraviolet  and infrared cutoffs and impose
the conditions not only on the dependence of the  Hamiltonian  on  the
cutoffs (as  usual)  but also on the dependence of the initial states.
It happens that one should consider the initial states to be singulary
dependent on   the   cutoffs   in  order  to  avoid  the  Stueckelberg
divergences. Different  types  of  semiclassical   approximations   to
quantum theory are discussed. It happens that the method of quantizing
classical solutions  to  field  equations  corresponds  not   to   the
WKB-approach but  to  the complex-WKB theory.  The problem of imposing
conditions on the semiclassical initial states is discussed. Different
prescriptions for choice of initial conditions are analysed.
}

\end{flushright}

\newpage

\section{Introduction}

There are  two  different languages to construct quantum field theory.
One of them is based on the notion of the $S$-matrix, another uses the
formalism of equations of motion.

These approaches  are  somehow  equivalent.  One  can  start  from the
$S$-matrix and obtain the Green functions \cite{BS} which determine in
the unique  fashion  all  the  properties  of the quantum field theory
according to  the   well-known   Whightman   reconstruction   theorem
\cite{axioms}. On  the other hand,  the scattering theory also can be
constructed as a corollary of the equations of motion.

However, there are several advantages and  disadvantages  of  each  of
formulations of   quantum  field  theory.  The  $S$-matrix  theory  is
manifestly Poincare-invariant  and  can   be   renormalized   by   the
procedure of the Bogoliubov-Parasiuk $R$-operation \cite{BS,Z}.  On the
other hand,  to  describe  time  evolution,  one  needs  a   dynamical
formalism based  on  the  equations  of  motion.  It is also much more
convenient to apply non-perturbative  methods  such  as  semiclassical
technique \cite{M1,M3} to the equations rather than to the $S$-matrix.

This talk  deals  with  constructing  the  semiclassical  field theory
within the  Hamiltonian  framework.  We  discuss  in  section  2   the
additional difficulties    arising    in    the   Schrodinger-equation
conception, as  well  as  possible  methods  to  resolve  them.  These
approaches will  be  very  useful  in  considering  the  corresponding
difficulties in the semiclassical theory. Section 3 deals with a brief
review of semiclassical methods in quantum mechanics.  In section 4 we
apply one of the semiclassical methods, the complex-WKB theory, to the
scalar quantum  field  model.  We  analyze the arising divergences and
renormalization. Section 5 contains concluding remarks.

\section{Hamiltonian field theory and its renormalization}

\subsection{The Haag theorem and Stueckelberg divergences}

{\bf 1}.
The main  disadvantage of the formalism of equations of motion is that
the structure of divergences is very complicated.  For example,  there
is a Haag theorem (see, for example, \cite{axioms}) that tells us that
the interaction representation cannot  be  defined  in  quantum  field
theory if the interaction is non-zero.

To illustrate  this  theorem,  consider  the  simple example of a free
scalar field of  mass  $m$.  This  theory  has  no  difficulties  with
divergences. However, if we consider the case of small mass and try to
divide the full Hamiltonian
$$
H= \int d{\bf x} \left[ \frac{1}{2}\pi^2({\bf x}) +
\frac{1}{2} (\nabla \varphi)^2 ({\bf x}) + \frac{m^2}{2}\varphi^2({\bf
x})\right]
$$
($\varphi$ is a field,  $\pi$ is a momentum) into two parts,  ``free''
Hamiltonian
$$
H_0 =
 \int d{\bf x} \left[ \frac{1}{2}\pi^2({\bf x}) +
\frac{1}{2} (\nabla \varphi)^2 ({\bf x})\right]
$$
and interaction Hamiltonian
$$
H_1=
\int d{\bf x} \frac{m^2}{2}\varphi^2({\bf x}),
\l{1}
$$
we would  find  that  each  of  two terms is not well-defined although
their sum is well-defined.  Namely, let us apply the operator $H_1$ to
the vacuum. The corresponding vector
$$
H_1|0> = \int \frac{d{\bf k}}{2\sqrt{{\bf k}^2+m^2}}
a_{\bf k}^+ a_{-{\bf k}}^+|0>
\l{2}
$$
expressing through  the  vacuum  state  $|0>$  and   operators
$a_{\bf k}^+$ creating a particle with momentum  ${\bf  k}$  does  not
belong to the Fock space. Namely, eq.\r{2} determines the two-particle
state with the wave function
$$
\Phi_2({\bf k},{\bf p})=\frac{1}{2\sqrt{{\bf k}^2+m^2}}
\delta({\bf k}+{\bf p})
$$
containing two singularities.  First of all, there is a delta-function
singularity corresponding to the infinite volume.  This is an infrared
divergence. Next, the full probability $\int d{\bf k} d{\bf p}
|\Phi_2({\bf k},{\bf p})|^2$ is also divergent because the multiplier
$\frac{1}{2\sqrt{{\bf k}^2+m^2}}$ does not sufficiently rapidly decrease
at the infinity. This corresponds to the ultraviolet divergence.

{\bf 2}.
Another difficulty of the  Hamiltonian  approach  is  the  problem  of
Stueckelberg divergences \cite{St}
(see also \cite{BS}). Let us consider the virtual process
of emission of the photon by the  electron.  When  one  considers  the
$S$-matrix, i.e.  the  infinite  time  interval,  such  a  process  is
forbidden by the  conservation  laws.  However,  at  the  finite  time
intervals this process takes place.  Consider the initial one-electron
state. The first-order perturbative evolution  operator  contains  the
term that transforms this state to the state containing two particles,
electron and photon.  The amplitude $\Phi_{{\bf p}{\bf k}}$ that the
momentum of the photon is
${\bf k}$  and  the  momentum  of  the  electron  is ${\bf q}$ has the
following asymptotic behaviour at ${\bf k},{\bf p} \to \infty$,  ${\bf
k}+{\bf p}=const$:
$$
\Phi_{{\bf p}{\bf k}}\sim |{\bf k}|^{-3/2}.
$$
Thus, the full probability to emit the virtual photon diverges as
$\int \frac{d{\bf k}}{|{\bf k}|^3}$ at larges ${\bf k}$.

Certainly, one  can  consider  the more physical processes as decay of
the $Z$-boson into a lepton-antilepton pair.  The same difficulty will
arise.

\subsection{Regularization and renormalization of the equation of motion}

{\bf 1}.
How should  one  eliminate such additional divergences?  First of all,
because of the Haag theorem it is necessary to  consider
not the  local quantum field theory interaction Hamiltonian of
the type
$$
H_1= \int d{\bf x} \varphi^n({\bf x})
$$
but its nonlocal analog (cf.\cite{reg})
$$
H_1^{\Lambda,L} = \int d{\bf x} g_L({\bf x}) \varphi_{\Lambda}^n({\bf x})
$$
expressed via the ultraviolet-cutoffed field
$$
\varphi_{\Lambda}({\bf x})= \int d{\bf y} A_{\Lambda}({\bf x}-{\bf y})
\varphi({\bf y}).
$$
When the  ultraviolet  and  infrared  cutoffs  $\Lambda$ and $L$ go to
infinity, the functions $g_L$ and $A_{\Lambda}$ behave as
$$
A_{\Lambda}({\bf x})\to \delta({\bf x}),
g_L({\bf x}) \to 1.
$$

{\bf 2}.
Consider the problem of renormalizing the Schrodinger equation
$$
i\dot{\Psi}_{\Lambda,L} = \hat{H}^{\Lambda,L}{\Psi}_{\Lambda,L}.
\l{3*}
$$
First of all, the quantum Hamiltonian $\hat{H}^{\Lambda,L}$ should not
have the simple form $H_0+H_1^{\Lambda,L}$.  It is necessary to add to
this expression  the  counterterms  in  order  to  make the $S$-matrix
finite. This means that  one  should  impose  the  conditions  on  the
dependence of  the  Hamiltonian on the cutoff parameters.  This is the
usual renormalization of the loop graphs \cite{BS}.

{\bf 3}.
But this renormalization procedure will not help us in eliminating the
Stueckelberg divergences  because  they arise at the level of the tree
Feynman graphs.  One can note that the  reason  for  the  Stueckelberg
divergences was  that  one  considered  ``bare''  electron  instead of
``physical'' electron consisting  of  electron  and  virtual  photons.
However, the  state of physical electron may singularily depend on the
cutoff parameters within the Hamiltonian framework  \cite{reg,F,A}. One
can hope   that  there  exists  an  unitary  operator  $T_{\Lambda,L}$
transforming bare particles to physical particles. This means that one
must consider  only  the  states  $\Psi_{\Lambda,L}$  depending on the
cutoff parameters $\Lambda,L$ as follows,
$$
\Psi_{\Lambda,L}=T_{\Lambda,L}\Phi.
$$
All the singularities of the amplitude $\Psi_{\Lambda,L}$ are involved
in the operator $T_{\Lambda,L}$, so that one can describe the physical
state by the regular state vector $\Phi$ rather than by  the  singular
vector $\Psi_{\Lambda,L}$.  The  Schrodinger  equation for the regular
vector $\Phi$ has the form:
$$
i\dot{\Phi} = T_{\Lambda,L}^+ H_{\Lambda,L} T_{\Lambda,L} \Phi.
\l{3}
$$
This means that we are considering another, non-Fock representation of
the canonical commutation relations (cf.\cite{nonFock,GM,B}).

\subsection{The Faddeev-type transformation}

Let us try to choose the transformation $T_{\Lambda,L}$ in a  suitable
way to  eliminate  the  Stueckelberg  divergences.  One  can  use  the
perturbative approach and seek for the transformation  $T_{\Lambda,L}$
in the following form \cite{F}:
$$
T_{\Lambda,L} = \exp [gA^1_{\Lambda,L}+g^2A_{\Lambda,L}^2+...]
\l{i2}
$$
for the anti-Hermitian operators $A^1_{\Lambda,L}$, $A^2_{\Lambda,L}$,
$...$ .  In the leading non-trivial order of perturbation theory,  one
obtains the  following form of the operator entering to the right-hand
side of eq.\r{3}:
$$
T^+_{\Lambda,L}H_{\Lambda,L}T_{\Lambda,L} =
H_0 + g(H_1^{\Lambda,L}- [A^1_{\Lambda,L},H_0])+...
$$
instead of
$$
H_{\Lambda,L} = H_0+gH_1^{\Lambda,L}+...
$$
This interaction  Hamiltonian can be chosen to be regular.  Of course,
the choice of the operator $A^1_{\Lambda,L}$ is not  unique.  However,
this ambiguity  corresponds  to the possibility of choice of different
(equivalent) representations of the commutation  canonical  relations.
All physical   quantities  such  as  $S$-matrix  or  decay  rates  are
invariant under change of the operators $A_{\Lambda,L}^k$ entering  to
the formula for the Faddeev-type transformation \r{i2}.

\subsection{The Bogoliubov $S$-matrix}

There is   also  another  approach  to  construct  the  transformation
$T_{\Lambda,L}$ which  is  based  on  the  Bogoliubov   procedure   of
switching on  the interaction
\cite{BS,switch}.  Let us investigate
for the simplicity the case of finite volume only.
Consider  the  interaction Hamiltonian
$$
H_1(t)= \xi(t) H_1,
\l{i1}
$$
instead of $H_1$,  where $\xi(t)$ is a time-dependent intensity of the
interaction. It  is  well-known  that  for  smooth  rapidly damping at
$\pm\infty$ function $\xi(t)$ the  corresponding  $S$-matrix  $S[\xi]$
can be renormalized \cite{BS}.

Consider the   function  $\xi_0$  depicted  on  fig.1.  This  function
corresponds to the interaction which is slowly swtched on and  rapidly
switched off.  It  happens that the transformation $T_{\Lambda,L}$ can
be chosen as
$$
T_{\Lambda,L} = S[\xi_0].
\l{4a}
$$

Let us  show  that  the  evolution  operator  corresponding   to   the
Hamiltonian entering to eq.\r{3},
$$
S^+[\xi_0] U_t S[\xi_0],
\l{4}
$$
where $U_t$ is the evolution operator in the original  theory  \r{3*},
is regular as $\Lambda,L\to \infty$.

Let us investigate some properties of the $S$-matrix.

1. For the intensity function
$$
\xi_t(\tau)= \xi_0(\tau-t)
$$
one has
$$
S[\xi_t] = e^{iH_0t} S[\xi_0] e^{-iH_0t}.
$$

2. For the function
$$
\tilde{\xi}_t(\tau) =
\left\{
\begin{array}{c}
\xi_0(\tau), \tau<0 \\
1, 0< \tau < t\\
0, \tau >t
\end{array}
\right.
$$
one has
$$
S[\tilde{\xi}_t] = e^{iH_0t} U_t S[\xi_0].
$$
Consider the following functions of switching the interaction
$$
\tilde{\xi}_1 = \tilde{\xi}_t, \qquad \tilde{\xi}_2 = \xi_t
$$
and corresponding smooth functions
$$
{\xi}_1(\tau) =
\left\{
\begin{array}{c}
\tilde{\xi}_1(\tau), \tau<t \\
\xi_0(t-\tau), t< \tau
\end{array}
\right.
\qquad
{\xi}_2(\tau) =
\left\{
\begin{array}{c}
\tilde{\xi}_2(\tau), \tau<t \\
\xi_0(t-\tau), t< \tau
\end{array}
\right.
$$
Since $\xi_1=\xi_2$  at  $\tau>t$,  the  product  $S^+[\xi_2]S[\xi_1]$
being regular  (because  $\xi_1,\xi_2$  are  smooth)   is   equal   to
$S^+[\tilde{\xi}_2]S[\tilde{\xi_1}]$. Making  use of the properties of
the Bogoliubov $S$-matrix, one obtains
$$
S^+[\tilde{\xi}_2] S[\tilde{\xi}_1] = e^{iH_0t}S^+[\xi_0]U_tS[\xi_0].
$$
We see that  the  operator  \r{4}  is  expressed  via  the  Bogoliubov
$S$-matrices $S[\xi_1]$  and  $S[\xi_2]$  corresponding  to the smooth
functions $\xi_1$ and $\xi_2$ and the evolution operator for the  free
field theory.   This   means  that  expression  \r{4}  is  regular  as
$\Lambda,L\to \infty$,  so that the operator  $T_{\Lambda,L}$  can  be
chosen as \r{4a}.

\section{Semiclassical methods}

\subsection{What is semiclassical approximation?}

Let us consider the  application  of  semiclassical  approximation  to
quantum field theory. First of all, it is necessary to clarify what is
semiclassical approximation.  In  quantum  mechanics  the  notion   of
semiclassical theory is usually associated with the WKB-approximation.
However, by now it seems to be not correct because there are different
semiclassical approaches and only one of them is the WKB-theory.

We consider  as semiclassical any method that allows us to investigate
the properties of solutions to the equation
$$
i\hbar \frac{\partial \psi({\bf x},t)}{\partial t}
= H \left(
{\bf x}, -i\hbar\frac{\partial}{\partial {\bf x}}
\right)
\psi({\bf x},t)
\l{s1}
$$
as $\hbar\to 0$. The small parameter $\hbar$ entering to eq.\r{s1} may
be a Planck constant, but may have no relationship with it. Equation of
the type \r{s1} may arise not only in quantum mechanical problems,  so
that semiclassical methods may be applicable  to  the  wide  class  of
physical problems. The only essential feature is that the coefficients
of the derivation operators are small (i.e.  proportional to the small
parameter $\hbar$),   while  the  coefficients  of  the  operators  of
multiplication in $x$ are of order $O(1)$.

\subsection{Classification of semiclassical methods}

Let us  present  (not  complete)   classification   of   semiclassical
approaches. First  of  all,  one can be interested in the structure of
the wave function $\psi$  or  in  the  average  values  expressed  via
$\psi$. An  example  of  the approach of the second type is the method
based on the Ehrenfest theorem.  The approaches of the first type  are
to be  presented.  It  should  be  noted  that  in  order to perform a
mathematical justification of the  semiclassical  asymptotics
\cite{M1},  it  is
more convenient  to consider the wave function rather than the average
values.

\subsubsection{Additive and multiplicative asymptotics}

One can  formulate  different  problems  concerning  the  asymptotical
properties of  the  wave  function  $\psi$.  For  example,  one can be
interested in small absolute error of the asymptotics,  i.e.  one  can
look for such a wave function $\psi_{as}$ that
$$
\psi - \psi_{as} = O(\hbar).
\l{s2}
$$
Otherwise, one can look for such asymptotical wave function  that  the
ralative error is small,
$$
\frac{\psi}{\psi_{as}} = 1+O(\hbar).
\l{s3}
$$
The problems of constructing asymptotics obeying  eq.\r{s2}  (additive
asymptotics) and eq.\r{s3} (multiplicative asymptotics) are different.
For example,  the multiplicative asymptotics for the ground state wave
function in a potential well is
$$
\psi_{as}(x) =\varphi(x) e^{-\frac{1}{\hbar}S(x)}.
\l{s4}
$$
Construction of additive asymptotics is a more simple  problem.  There
is a remarkable relation \cite{M3}
$$
e^{-\frac{x^2+x^4}{\hbar}} = e^{-\frac{x^2}{\hbar}} + O(\hbar)
$$
which shows  that  one  can consider the quadratic function instead of
$S$ in  eq.\r{s4}.  This  corresponds  to  the  substitution  of   the
arbitrary potential  by  the  oscillator  potential.  We  see that the
oscillator approximation is one of  the  types  of  the  semiclassical
approximation.

\subsubsection{WKB and complex-WKB methods}

There are   different   additive   asymptotics  abeying  approximately
eq.\r{s1}. For example, the WKB wave function
$$
\psi_{as}({\bf x},t)  =  \varphi({\bf  x},t)  e^{\frac{i}{\hbar}S({\bf
x},t)}
\l{s5}
$$
is an asymptotic solution to eq.\r{s1}.

On the other hand, the wave function
$$
\psi_{as} = const
e^{\frac{i}{\hbar}S(t)}
e^{\frac{i}{\hbar}({\bf P}(t)({\bf x}-{\bf Q}(t)))}
f\left(t, \frac{{\bf x}-{\bf Q}(t)}{\sqrt{\hbar}}\right)
\l{s6}
$$
also approximately obeys eq.\r{s1}, although it is not of the WKB-type.
Thus, the  WKB-method  is  not  the  only  semiclassical  method.  The
technique that allows us to construct wave functions  like  \r{s6}  is
called as the complex-WKB method \cite{M3}.

There is  a  much  more general semiclassical approach,  the theory of
Lagrangian manifolds with complex germs
\cite{M3}.  This  theory  allows  us  to
construct $n$-dimensional  wave  functions  that are not exponentially
small at the small vicinity of the $k$-dimensional surface.  As $k=n$,
this theory  corresponds  to  the  WKB theory,  while the case
$k=0$ is associated with the complex-WKB method.

\subsection{Applications to quantum field theory}

Different quantum mechanical approaches  can  be  generalized  to  the
quantum field   theory.   For   example,   the   analog  of  tunneling
approximation (corresponding  to  the  problem  of  constructing   the
multiplicative asymptotics)  is  the theory of instantons
\cite{inst} and bounces \cite{bounce},
while the  complex-WKB  theory  is  relevent  to  the  static  soliton
quantization approach
\cite{sol},
nonequilibrium field theory \cite{noneq}, physics of strong
electromagnetic and gravitational fields
\cite{str}.  The  theory  of  Lagrangian
manifolds with  complex  germs  is  associated  with  the  problems of
considering the systems with  integrals  of  motion,  the  constrained
systems, interpretation   of   soliton  zero  modes,  quantization  of
periodic solutions (see, for example, \cite{R}).

\section{Semiclassical field theory and its renormalization}

\subsection{Formal semiclassical theory}

Consider the problem of semiclassical approximation  for  the  quantum
field theory.  For the simplicity,  investigate the case of the scalar
theory with the Lagrangian
$$
{\cal L} = \frac{1}{2}\partial_{\mu}\varphi \partial_{\mu} \varphi
-\frac{1}{g}V(\sqrt{g}\varphi) - V_1(\sqrt{g}\varphi) -...,
$$
where the  term  $\frac{1}{g}V(\sqrt{g}\varphi)$  corresponds  to  the
classical Lagrangian,  while $V_1$ is a one-loop counterterm. From the
Hamiltonian point  of  view,  one can write the functional Schrodinger
equation,
$$
i\dot{\Psi}[t,\varphi(\cdot)] = \int d{\bf x}
\left[ -\frac{1}{2}\frac{\delta^2}{\delta    \varphi({\bf    x})\delta
\varphi({\bf x})}
+\frac{1}{2}(\nabla\varphi)^2({\bf x}) +
\frac{1}{g} V(\sqrt{g}\varphi) + V_1(\sqrt{g}\varphi)\right]
\Psi[t,\varphi(\cdot)]
\l{q1}
$$
for the functional $\Psi[t,\varphi(\cdot)]$. After rescaling
$$
\sqrt{g}\varphi=\tilde{\varphi}
$$
and multiplying by $g$ eq.\r{q1} is taken to the  semiclassical  form,
since the  coefficients  of  all  derivation  operators  are  of order
$O(g)$, while the coefficients of the operator  of  multiplication  be
$\tilde{\varphi}$ are of order $O(1)$.

Let us  apply  the  complex-WKB  technique  to  eq.\r{q1}.  One should
consider the following wave functional analogous to \r{s6}:
$$
\Psi[t,\varphi(\cdot)] = e^{\frac{i}{g}S^t}
e^{\frac{i}{\sqrt{g}}\int d{\bf x}\Pi^t({\bf x})
(\varphi({\bf x})- \frac{\Phi^t({\bf x})}{\sqrt{g}})}
f^t\left(\varphi(\cdot)-\frac{\Phi^t(\cdot)}{\sqrt{g}}\right).
\l{q2}
$$
Substituting expression \r{q2} to eq.\r{q1} and making equal the terms
of order $O(1/g)$, one obtains the condition on the phase,
$$
S^t=\int_0^t dt d{\bf x} {\cal L}.
$$
The order $O(g^{-1/2})$ gives us the classical equations of motion,
$$
\dot{\Pi}^t=\Phi^t, \qquad \partial_{\mu}\partial_{\mu}\Phi^t
+V'(\Phi^t)=0,
$$
while the order $O(1)$ corresponds to the Schrodinger-type
equation on $f$
$$
i\dot{f}=H_2 f
\l{q3}
$$
with the quadratic Hamiltonian
$$
H_2 = \int d{\bf x}
\left[
-\frac{1}{2}\frac{\delta^2}{\delta \phi({\bf x})\delta \phi({\bf x})}
+\frac{1}{2} (\nabla \phi)^2({\bf x})
+\frac{1}{2}V''(\Phi^t({\bf x}))\phi^2({\bf x})
+V_1(\Phi^t({\bf x}))
\right].
\l{q4}
$$
One can notice that  the  complex-WKB  method  is  equivalent  to  the
procedure of extracting the classical field component,
$$
\varphi = \frac{\Phi}{\sqrt{g}} + \phi
$$
and quantizing the obtained theory.

One can construct then the exact solutions to eq.\r{q3}.  The simplest
way is to consider the Gaussian ansatz
$$
f^t[\phi(\cdot)] =
c^t \exp
\left[
\frac{i}{2} \int  d{\bf  x}d{\bf  y}  \phi({\bf  x})  \tilde{R}^t({\bf
x},{\bf y}) \phi({\bf y})
\right]
\l{q4*}
$$
which obeys eq.\r{q3} if
$$
\begin{array}{c}
\dot{\hat{R}}^t + \hat{R}^t \hat{R}^t +
(-\Delta +  V''(\Phi^t({\bf x})))=0,\\
i\dot{c}^t =
(-\frac{i}{2}Tr \hat{R}^t + \int d{\bf x} V_1(\Phi^t({\bf x})))
c^t,
\end{array}
\l{q5}
$$
where $\hat{R}^t$ is the operator with kernel $\tilde{R}^t$,
$$
(\hat{R}^tf)({\bf x})= \int d{\bf y} \tilde{R}^t({\bf x},{\bf y})
f({\bf y}).
$$
We see that  the  Gaussian  approximation  considered  as  variational
approximation in \cite{G} is exact as $g\to 0$.

The more complicated ansatz for eq.\r{q3} is the following,
$$
\Lambda^+[\delta\Phi^t_1]
...
\Lambda^+[\delta\Phi^t_k]
f^t[\phi(\cdot)]
$$
where operators
$$
\Lambda^+[\delta \Phi] =
\int d{\bf x}
[\frac{d}{dt}\delta\Phi^t \phi({\bf x})
-\delta\Phi^t({\bf x}) \frac{1}{i}\frac{\delta}{\delta \phi({\bf x})}]
$$
commute with $i\frac{d}{dt}-H_2$ and transform a solution to eq.\r{q3}
to another solution if
$$
\partial_{\mu}\partial_{\mu} \delta\Phi^t + V''(\Phi^t)
\delta \Phi^t = 0.
$$
We see that the function $\delta\Phi^t$ plays a role of a  fluctuation
about the classical solution.

The problem  of  renormalization  of eq.\r{q5} arises.  To investigate
this problem, it is convenient to use the operational technique
\cite{M2}.

\subsection{Operational calculus}

To analyse eq.\r{q5},  it is convenient to present all  the  operators
via the differentiation operator $-i\frac{\partial}{\partial {\bf x}}$
and the operator of multiplication by ${\bf x}$. The function
$$
A({\bf x},{\bf k})
$$
such that
$$
\hat{A} = A({\bf x},-i\frac{\partial}{\partial {\bf x}})
$$
is called as a symbol of the  operator  $\hat{A}$
\cite{M2}.  The  procedure  of
reproducing the  operator  $\hat{A}$  from  the  symbol  $A$ is called
quantization. The main difficulty is  that  numbers  $x_i$  and  $k_i$
commute, while  operators  $x_i$  and $-i\partial/\partial x_i$ do not
commute. This means that one should specify the ordering of coordinate
and momentum  operators  and  obtain  then  different  presciptions of
quantization. For example, the $qp$-quantization is the following: one
puts the  differentiation  operators  to  the  right,  the  coordinate
operators to the left, so that the matrix element
$
<{\bf x}|\hat{A}|{\bf k}>
$
of the operator $\hat{A}$ between states $<{\bf x}|$ and  $|{\bf  k}>$
with given   coordinate   ${\bf  x}$  and  given  momentum  ${\bf  k}$
correspondingly is expressed via its symbol
$$
<{\bf x}|\hat{A}|{\bf k}> = A({\bf x},{\bf k}) <{\bf x}|{\bf k}>.
\l{q6}
$$
Eq.\r{q6} can be viewed as a definition of the $qp$-quantization,
$$
\hat{A}=A
(\stackrel{2}{\bf x},
\stackrel{1}{-i\frac{\partial}{\partial {\bf x}}}).
$$
All operational calculus can be reformulated in terms of symbols.  For
example, let  $A$  and  $B$ be $qp$-symbols of the operators $\hat{A}$
and $\hat{B}$,
$$
\hat{A}=A
(\stackrel{2}{\bf x},
\stackrel{1}{-i\frac{\partial}{\partial {\bf x}}}),
\qquad
\hat{B}=B
(\stackrel{2}{\bf x},
\stackrel{1}{-i\frac{\partial}{\partial {\bf x}}}).
$$
Then the sum of these operators $\hat{A}+\hat{B}$ corresponds  to  the
sum of the sumbols $A+B$, while the symbol of the product
$$
\hat{A}\hat{B} = (A*B)
(\stackrel{2}{\bf x},
\stackrel{1}{-i\frac{\partial}{\partial {\bf x}}})
$$
is
$$
(A*B)({\bf x},{\bf k}) =
A
(\stackrel{2}{\bf x},
\stackrel{1}{{\bf k}-i\frac{\partial}{\partial {\bf x}}})
B({\bf x},{\bf k}).
\l{q7}
$$
Eq.\r{q7} can be viewed as a quantum ``multiplication'' of functions.

Eq.\r{q7} is very suitable to analyze the asymptotic behaviour of  the
symbol of the operator $\hat{A}\hat{B}$ at large values of $|{\bf k}|$.
Namely, expression \r{q7} can be  presented  as  a  formal  asymptotic
series
$$
(A*B)({\bf x},{\bf k}) =
A({\bf x},{\bf k}) B({\bf x},{\bf k}) -
i\frac{\partial A}{\partial {\bf k}}
\frac{\partial B}{\partial {\bf x}}
-\frac{1}{2}
\frac{\partial^2A}{\partial k_m \partial k_n}
\frac{\partial^2B}{\partial x_m \partial x_n}
+ ...
\l{q8}
$$
with the  first  term  being  a  classical  product.  Eq.\r{q8}  is an
expansion in $1/|{\bf k}|$ as $|{\bf k}| \to \infty$.  Namely,  if  $A
\sim |{\bf  k}|^{-a}$,  $B \sim |{\bf k}|^{-b}$ as ${\bf k}\to\infty$,
the first term is of order $|{\bf k}|^{-a-b}$, the next being of order
$|{\bf k}|^{-a-b-1}$  etc.  It is very important that $A$ and $B$
should have
no oscillating factors like $e^{i|{\bf k}|S}$ at large $|{\bf k}|$.

The trace of the operator can be also expressed via its symbol,
$$
Tr \hat{A} = \frac{1}{(2\pi)^d}
\int d{\bf x}d{\bf k} A({\bf x},{\bf k}),
\l{q9}
$$
so that one can investigate the divergences in $Tr \hat{R}$  entering
to eq.\r{q5} with the help of eq.\r{q9}.

\subsection{Possible Schrodinger functionals. Free theory}

Not arbitrary  operator  $\hat{R}^t$  can  enter  to the Gaussian wave
functional \r{q4*}.  First of all, let us investigate this problem for
the simple  case  of  free  theory.  The  Gaussian  functional \r{q4*}
corresponds to the Gaussian Fock space vector
$$
\exp[\frac{1}{2} \int d{\bf x}d{\bf y} a^+({\bf x})
\tilde{B}({\bf x},{\bf y}) a^+({\bf y})] |0>,
\l{q10}
$$
where
$$
a^{\pm}({\bf x}) = \int \frac{d{\bf k}}{(2\pi)^{d/2}}
a^{\pm}_{\bf k} e^{\mp i{\bf k}{\bf x}}
$$
are linear  combinations  of creation (annihilation) operators.  It is
well-known \cite{B}
that eq.\r{q10} corresponds to the Fock space state if
$$
||\hat{B}||<1, \qquad Tr \hat{B}^+\hat{B} < \infty.
$$
This means that the following conditions  should  be  imposed  on  the
asymptotical behaviour of the symbol of the operator $\hat{B}$,
$$
B({\bf x},{\bf k}) \sim \frac{1}{|{\bf k}|^{d/2+\varepsilon}},
\qquad |{\bf k}|\to\infty.
\l{q11}
$$
The Gaussian  quadratic  form  entering to eqs.\r{q4*} and \r{q10} are
associated each other,
$$
\hat{R} = i\sqrt{\hat{\omega}} (1-\hat{B})(1+\hat{B})^{-1}
\sqrt{\hat{\omega}},
\l{q12}
$$
where $\hat{\omega}$ is the operator $\hat{\omega}=\sqrt{-\Delta+m^2}$
with the  symbol  $\omega_{\bf  k}$.  It  follows from eqs.\r{q11} and
\r{q12} that the symbol of the operator $R$ should have the  following
asymptotical behaviour as $|{\bf k}|\to\infty$,
$$
R({\bf x},{\bf k})=
i\omega_{\bf k} + O(|{\bf k}|^{-(d/2+\varepsilon-1)}).
\l{q13}
$$

\subsection{Possible Schrodinger functionals. General case}

Let us impose the condition on the Gaussian functionals \r{q4*}
\cite{MS97}. First
of all,  notice that the condition \r{q13} is not invariant under time
evolution. Namely,  consider the equation on  the  operator  $\hat{R}$
\r{q5} which can be presented as the equation on its $qp$-symbol $R$:
$$
\dot{R^t}+ R^t * R^t + \omega_{\bf k}^2 + f_t =0,
\l{q14}
$$
where
$$
f_t({\bf x},{\bf k}) = f_t({\bf x}) =
V_{int}'' (\Phi^t({\bf x}))= V''(\Phi^t({\bf x}))-m^2,
\quad
\omega_{\bf k}=\sqrt{{\bf k}^2+m^2}.
$$
One can expand the symbol $R^t$ in $1/|{\bf k}|$,
$$
R = i\omega_{\bf k} + R_1 +...
$$
where $R_1$ obeys the equation,
$$
\dot{R}_1 + 2i\omega_{\bf k}R_1 + f_t({\bf x})
= O(|{\bf k}|^{-1}).
$$
the solution to this equation is expressed as
$$
R^t_1 ({\bf x},{\bf k}) =
\frac{if_t({\bf x})}{2\omega_{\bf k}}
+ \frac{\varphi_t({\bf x},{\bf k}/\omega_{\bf k})}
{\omega_{\bf k}} e^{-2i\omega_{\bf k}t}
$$
for some  $\varphi_t$.  We  see  that  the condition \r{q13} should be
improved at $d+1\ge 5$.

One can also notice that the conditions
$$
\begin{array}{c}
R({\bf x},{\bf k}) = i\omega_{\bf k}
+ \frac{i}{2\omega_{\bf k}} V_{int}''(\Phi({\bf x}))
+O(\frac{1}{|{\bf k}|^{d/2-1+\varepsilon}}),
\quad d=5,6\\
R({\bf x},{\bf k}) = i\omega_{\bf k}
+ \frac{i}{2\omega_{\bf k}} V_{int}''(\Phi({\bf x}))
-
\frac{1}{4}
\left(
\frac{\partial}{\partial t}
+ \frac{{\bf k}}{\omega_{\bf k}}
\frac{\partial}{\partial {\bf x}}
\right)
\frac{V_{int}''(\Phi({\bf x})}{\omega_{\bf k}^2}
+O(\frac{1}{|{\bf k}|^{d/2-1+\varepsilon}}),
\quad d=7,8
\end{array}
\l{q15}
$$
are invariant under time evolution.

These conditions  can  be  interpretted  from the point of view of the
Bogoliubov procedure of switching the interaction. Let us consider the
theory with interaction \r{i1} with the  function  $\xi_0$  depicted  on
fig.1, One  can  apply the semiclassical complex-WKB technique to this
theory and obtain the equation on $R$:
$$
\dot{R^t}+ R^t * R^t + [\omega_{\bf k}^2 +
\xi_0(t) V_{int}''(\Phi({\bf x}))]
=0
$$
instead of \r{q14}.
One should also impose the conditions at $t<-T_1$:
$$
R^{-T_1}({\bf x},{\bf k}) = i\omega_{\bf k}
+
O(\frac{1}{|{\bf k}|^{d/2-1+\varepsilon}}).
$$
Applying a regular perturbation theory in $1/|{\bf k}|$, one finds the
conditions \r{q15}.

\section{Conclusions}
We conclude the talk by adding the following remarks.

1. Renormalization procedure is usually associated with  loop  Feynman
graphs. We  have  seen  that for time-dependent dynamical theory it is
necessary to consider the divergences for the tree  graphs,  i.e.  for
the classical field theory. The fact that there exist the Stueckelberg
divergences in the perturbation theory is shown to imply that the free
condition on the Gaussian quadratic form, eq.\r{q13}, is not inveriant
under time  evolution.  Non-triviality of
initial   conditions   for   the
perturbation theory which is  associated
with Faddeev  or Bogoliubov transformations leads to the nontriviality
of the initial conditions on the Gaussian quadratic form.

2. The invariance of the conditions \r{q15} under time evolution
is not mathematically proved in this talk. Hovever, the mathematical
formulation of this statement is presented in \cite{DAN}.

3. One  can compare the obtained conditions on the quadratic form $R$,
eqs.\r{q15}, with  known  conditions.  For  example,   free   initial
condition is  not invariant under time evolution.  The conjecture that
one should diagonalyse the Hamiltonian at each moment  of  time
\cite{GM} (this
corresponds to the condition
$\hat{R} \sim \sqrt{-\Delta+m^2+V_{int}''(\Phi)}
$)
is also correct only for sufficiently small dimensions.

4. One  can  investigate  also  the  pre-exponential  factor  in   the
Schrodinger wave  functional \cite{MS97}.  One can notice that some of
the divergences are eliminated  by  the  one-loop  counterterm  $V_1$,
while other divergences should be involved into the condition on $c$.
This means that the factor $c$ may be singular as  $\Lambda\to\infty$,
so that the Schrodinger representation is singular for this case.  One
can understand this with  the  help  of  the  following  analogy.  The
expression
$
\exp(\frac{1}{2}\sum_k c_k (a_k^+)^2)|0>
$
determines the Fock space vector if $c_k <1$ for all $k$ and
$\sum_k |c_k|^2< \infty$.  However, the Schrodinger representation for
this vector  exists if $\sum_k |c_k|<\infty$,  so that the Schrodinger
representation may  not exist even for the well-defined Fock vector.
This may be analogous to the quantum field theory.

5. One can generalize the presented approach to the  more  complicated
field systems. Generalization to the fermionic case is straightforward
since it is necessary to extract the  classical  parts  only  from  the
bosonic degrees   of  freedom.  To  generalize  the  approach  to  the
constrained systems,  one can perform the gauge fixing  procedure  and
apply this  approach.  One  can also perform a direct investigation of
the Hamiltonian formulation of the  gauge  theories  by  applying  the
theory of  Lagrangian  manifolds  with  complex  germs
\cite{M3}.  The developed
approach is also useful to investigate the large-N systems
(see \cite{MSh} for more details).

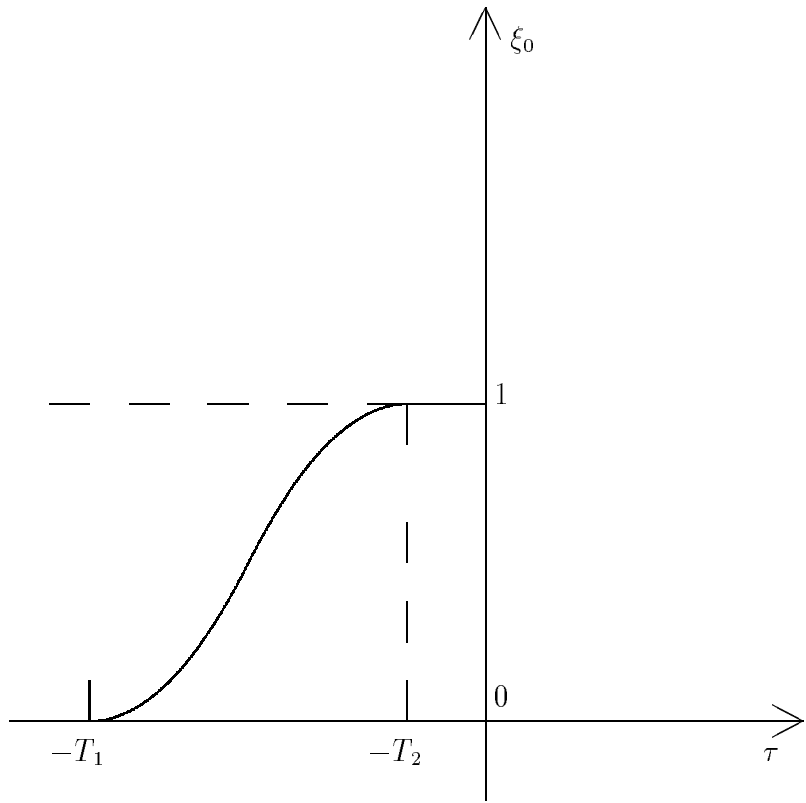
\end{document}